\documentclass[aps,nofootinbib,floatfix,showpacs,preprintnumbers]{revtex4} 
\usepackage{graphicx}
\input epsf

\def\gev{\,{\rm GeV}\,}
\def\lsim{\mathrel{\raise.3ex\hbox{$<$\kern-.75em\lower1ex\hbox{$\sim$}}}}
\def\gsim{\mathrel{\raise.3ex\hbox{$>$\kern-.75em\lower1ex\hbox{$\sim$}}}}
\def\gray{\,$\gamma$-ray\ }
\def\grays{\,$\gamma$-rays\ }

\def\sigv{$\langle\sigma v\rangle$ }

\def\cmm2{{\,\rm cm^{-2}}}
\def\cm2{{\,{\rm cm}^2}}
\def\cmm3{{\,{\rm cm}^{-3}}}
\def\gcmm3{{\,{\rm g\,cm^{-3}}}}

\def\fun#1#2{\lower3.6pt\vbox{\baselineskip0pt\lineskip.9pt
  \ialign{$\mathsurround=0pt#1\hfil##\hfil$\crcr#2\crcr\sim\crcr}}}

\def\be{\begin{equation}}
\def\ee{\end{equation}}
\def\bea{\begin{eqnarray}}
\def\eea{\end{eqnarray}}

\newcommand{\ec}[1]{Eq.~(\ref{eq:#1})}

\newcommand{\eql}[1]{\label{eq:#1}}

\newcommand{\fwimp}{$f_{\rm{WIMP}}$\,}

\begin{document}

\title{A Robust Approach to Constraining Dark Matter Properties with
  Gamma-Ray Data}

\author{Eric J. Baxter$^1$, Scott Dodelson$^{1,2,3}$}

\affiliation{$^1$Department of Astronomy \& Astrophysics, The
University of Chicago, Chicago, IL~~60637}
\affiliation{$^2$Center for Particle Astrophysics, Fermi National
Accelerator Laboratory, Batavia, IL~~60510}
\affiliation{$^3$Kavli Institute for Cosmological Physics, Chicago, IL~~60637}
\date{\today}

\begin{abstract}
Photons produced in the annihilations of dark matter particles can be
detected by gamma-ray telescopes; this technique of indirect detection
serves as a cornerstone of the upcoming assault on the dark matter
paradigm.  The main obstacle to the extraction of information about
dark matter from the annihilation photons is the presence of large and
uncertain gamma-ray backgrounds.  We present a new technique for using
gamma-ray data to constrain the properties of dark matter that makes
minimal assumptions about the dark matter and the backgrounds.  The
technique relies on two properties of the expected signal from
annihilations of the smooth dark matter component in our galaxy: 1) it
is approximately rotationally symmetric around the axis connecting us
to the Galactic Center, and 2) variations from the mean signal are
uncorrelated from one pixel to the next.  We apply this technique to
recent data from the Fermi telescope to generate constraints on the
dark matter mass and cross section for a variety of annihilation
channels.  We quantify the uncertainty introduced into our constraints
by uncertainties in the halo profile and by the possibility that the
halo is triaxial.  The resultant constraint, the flux $F\le 4.5\times
10^{-6}$ cm$^{-2}$ s$^{-1}$ sr$^{-1}$ for energies between 1 and 100
GeV at an angle $15^\circ$ away from the Galactic Center, translates
into an upper limit on the velocity weighted annihilation cross
section of order $10^{-25}$ cm$^3$ s$^{-1}$ depending on the
annihilation mode.
\end{abstract}
\maketitle

\section{Introduction}

Evidence for the existence of non-baryonic dark matter has been
accumulating for many decades.  Combined constraints from measurements
of anisotropies in the cosmic microwave background radiation, the
shape of the galaxy power spectrum, and the Hubble constant fix both
the total matter and the baryon densities, indicating that
non-baryonic dark matter makes up 85\% of the matter density of the
universe \cite{Jarosik:2010}.  Despite the preponderance of evidence
for its existence, however, little is known about the identity of the
dark matter.  One way to glean information about the properties of
this mysterious substance is through {\it indirect detection}, the
observation of the annihilation products of dark matter particles.

Indirect detection is an attractive prospect for several reasons.  Its
primary advantage is that, while dark matter
itself is very difficult to detect, the annihilation products of dark matter particles may
be easily detectable.  If photons are produced in dark matter
annihilations, for instance, existing telescopes can detect them.
Another attractive feature of indirect detection is that the
velocity-weighted, thermally averaged annihilation cross section of
the dark matter, $\left< \sigma v \right>$, which governs the expected
indirect detection signal, is constrained if the dark matter is a
thermal relic.  Additionally, indirect detection has the potential to
reveal information about the distribution of dark matter beyond our
local environment.  For these and other reasons, indirect detection
nicely compliments the other techniques that may be used to identify
dark matter: direct detection and collider searches
\cite{Bergstrom:2010}.  It is likely that all three techniques will be
necessary for a definitive identification of the dark matter.

The present is an exciting time for indirect detection as a number of
experiments are currently underway that are capable of detecting a
signal from dark matter annihilations.  Neutrino detectors such as
IceCube and AMANDA \cite{Landsman:2007}, air Cherenkov detectors such
as H.E.S.S. \cite{Horns:2008} and VERITAS \cite{Holder:2006}, cosmic
ray detectors such as PAMELA \cite{Picozza:2007}, and space based
\gray telescopes such as the Fermi Gamma-Ray Space Telescope (FGST)
\cite{Atwood:2009} are all poised to make important contributions to
the indirect detection of dark matter.

In this paper, we focus on the possibility of using \gray data taken
by the Large Area Telescope (LAT) on board the FGST to constrain the
properties of dark matter.  The LAT is a wide field, pair conversion
\gray telescope that covers an energy range from about 20 MeV to 300
GeV \cite{Atwood:2009}.  Gamma-rays are particularly well suited for
indirect detection because they are relatively easy to detect, they
are produced in many models of dark matter annihilation, and they
propagate through the universe with small optical depth (especially at
low energies).

Unfortunately, although \grays may in principle contain significant
information about dark matter, the process of extracting this
information is severely complicated by the presence of large and
uncertain backgrounds to the dark matter signal.  The primary
challenge of indirect detection using \grays is therefore to extract a
signal which may be hidden in backgrounds that are larger by orders of
magnitude.  Two primary features of the detected photons are their
energy and arrival direction. A number of studies have used these two
features to extract the dark matter signal from the background. For
example, Ref.~\cite{Abdo:2010dm} used only energy information, while
Ref.~\cite{Zaharijas:2010} and Ref. \cite{hooperandgoodenough} used
both the spectral and angular distribution information. Given the
photon counts, derived quantities can also be used to distinguish the
signal from the backgrounds. The probability distribution function
(PDF) has been proposed as a discriminant by a number of
groups~\cite{Lee:2009,Dodelson:2009ih,Baxter:2010}.  Anisotropy of the
distribution, especially when combined with spectral information, has
also been proposed as a powerful way of extracting the
signal~\cite{SiegalGaskins:2009ux,Hensley:2009gh}.

Here we propose a new and robust approach for constraining dark matter
from \gray data that uses only the angular distribution of the
photons.  We rely on two important aspects of the dark matter signal
to help separate it from backgrounds.  First, part of the expected
signal from dark matter in our galaxy is smooth (i.e. variations from
the mean flux in nearby pixels are uncorrelated).  Second, the dark
matter signal comes from a nearly spherically symmetric halo, so the
signal is azimuthally symmetric about the axis connecting us to the
center of our Galaxy. This is in sharp contrast to the backgrounds,
which are heavily concentrated near the disk of the Galaxy.  More
generally, astrophysical backgrounds have different morphologies and
may be clumped (i.e. variations from the mean flux may be correlated
in nearby pixels).  These differences between the signal and the
backgrounds allow us to remove some of the contribution from the
backgrounds to place an interesting limit on the dark matter.  The
approach, which we call a {\it Ring Analysis}, makes very minimal
assumptions about the nature of the signal and no assumptions about
the backgrounds.  Therefore, this approach is very conservative and
will lead to robust limits on the properties of the dark matter.

In \S\ref{sec:data} we describe the way we processed the Fermi LAT
data to generate photon count and exposure maps. In
\S\ref{sec:ringanalysis} we present the Ring Analysis technique that
we have developed for constraining the presence of an azimuthally
symmetric signal on the sky. Consider an annulus centered on the axis
connecting us to the Galactic Center, identified by the angle $\psi$
between this axis and the annulus. The Ring Analysis results in an
upper limit on any contribution to the flux that is uniform in this
annulus. Fig.~\ref{fig:ring_psi_plot} presents these upper limits on
the uniform flux from the Fermi LAT data. Transforming these upper
limits into constraints on the properties of dark matter particles
requires a number of steps and assumptions. These, and in particular
the uncertainties involved, are discussed in \S\ref{sec:application}.
Our conclusions are presented in \S\ref{sec:summary}.

\section{Data}
\label{sec:data}

The Fermi LAT is a pair conversion \gray detector that operates
roughly in the energy range from 20 MeV to 300 GeV.  A scintillating
anti-coincidence detector allows for the rejection of contaminating
high energy particle events.  The specifications of the detector are
described in detail in Ref. \cite{Atwood:2009}.  Our analysis is based on
LAT data downloaded in the form of weekly all-sky releases from the
Fermi Science Support Center website at
http://fermi.gsfc.nasa.gov/ssc/data/.

Even with the on-board anti-coincidence detector there is a residual
background of particles that are misclassified as \grays by the LAT
detectors.  This poses a challenge for our analysis because these
misinterpreted cosmic rays constitute a potentially large and
uncertain background.  The Fermi collaboration has made public the
DataClean event class which implements several data cuts to minimize
cosmic ray contamination as well as improved particle background and
instrument modeling.  Their data selection techniques and event
modeling are described in Ref.~\cite{Abdo:2010}.  We restrict our
analysis to only the events labeled as DataClean and use the
corresponding P6\_V3\_DATACLEAN instrument response function to
calculate exposure maps.

Following Ref.~\cite{Abdo:2010}, we confine our analysis to those events
coming from zenith angles $< 100^{\circ}$ in order to reduce
contamination by \grays produced in cosmic ray interactions with the
Earth's atmosphere.  The resulting data set amounts to an exposure of
roughly $7\times10^{10} \cm2 \rm{s}$ across the sky covering a date range
from 2008-08-04 to 2010-12-07.  

\begin{figure}[htpb]
\includegraphics[scale = 0.5,angle=90]{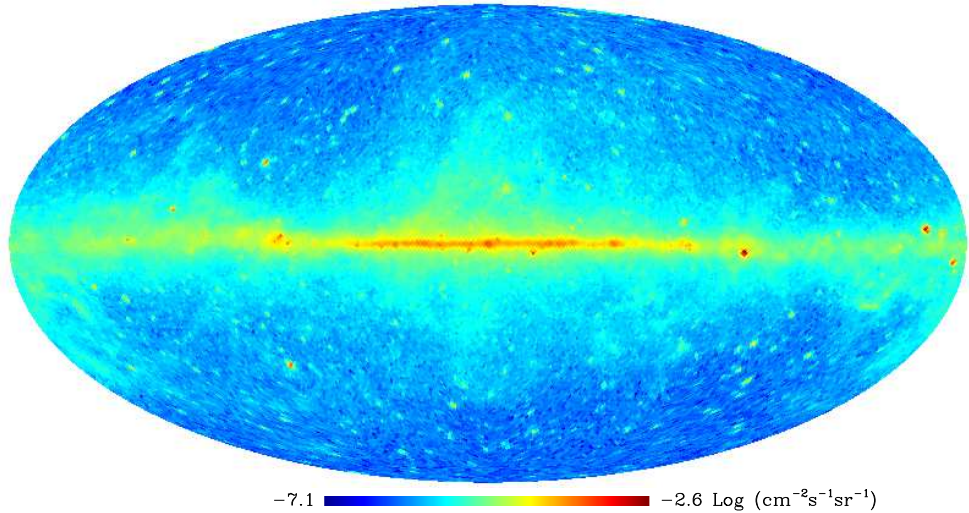}
\caption{All-sky map of the \gray flux between 1 GeV and 100 GeV as
  measured by the FGST with an exposure of roughly $7\times10^{10}
  \cm2 \rm{s}$.}
\label{fig:skymap}
\end{figure}

We produce photon count and exposure maps using the {\it GaDGET}
package, a set of software routines designed for use with the LAT data
by the Fermi collaboration \cite{Ackermann2008}.  In generating the
maps, the data were divided into 29 energy bins between 1 GeV and 100
GeV, logarithmically spaced in energy.  This restriction in energy was
chosen because the LAT performance is well characterized in this
energy range.  Because the generation of these maps is computationally
intensive, the data were processed in parallel on the
Fulla\footnote{http://fulla.fnal.gov/} computing cluster at Fermilab.
The maps generated using the {\it GaDGET} software were then converted
to the HEALPix\footnote{http://healpix.jpl.nasa.gov/} isolatitude
pixelization scheme with $N_{\rm{side}}=64$, corresponding to a pixel
size of roughly $(1^{\circ})^2$ (comparable to the width of the point
spread function (PSF) of Fermi at the lowest energy we consider).
Fig.~\ref{fig:skymap} shows the resulting all-sky \gray map in the energy
range $1 \gev < E < 100 \gev$.

\section{Ring Analysis}
\label{sec:ringanalysis}

\subsection{Overview}

The Ring Analysis technique that we develop in this paper is a method
for constraining the presence of a signal on the sky that satisfies
two requirements: 1) it is azimuthally symmetric, and 2) variations
from the mean signal at any zenith angle are uncorrelated from one
pixel to the next.  As we discuss in \S\ref{sec:application}, the
annihilation signal from smooth dark matter is expected to satisfy
these two requirements, and we can quantify the extent to which the signal deviates from these
conditions. In this section,
however, we make no reference to the nature of the signal itself.

A consequence of these two features is that the signal in each pixel
of a ring of constant $\psi$ (where $\psi$ is the zenith angle) can be
considered to be drawn independently from some underlying
distribution.  In the statistical literature, such a signal is
referred to as independent and identically distributed, or i.i.d.  The
constraint will be on the amplitude of any i.i.d. component of the
data.  In the case of interest, the backgrounds dwarf the signal and
the background flux in each pixel may depend on nearby pixels and may
vary as a function of the angle transverse to $\psi$, which we label
$\phi$.

\begin{figure}[htpb]
\includegraphics[scale = 0.7]{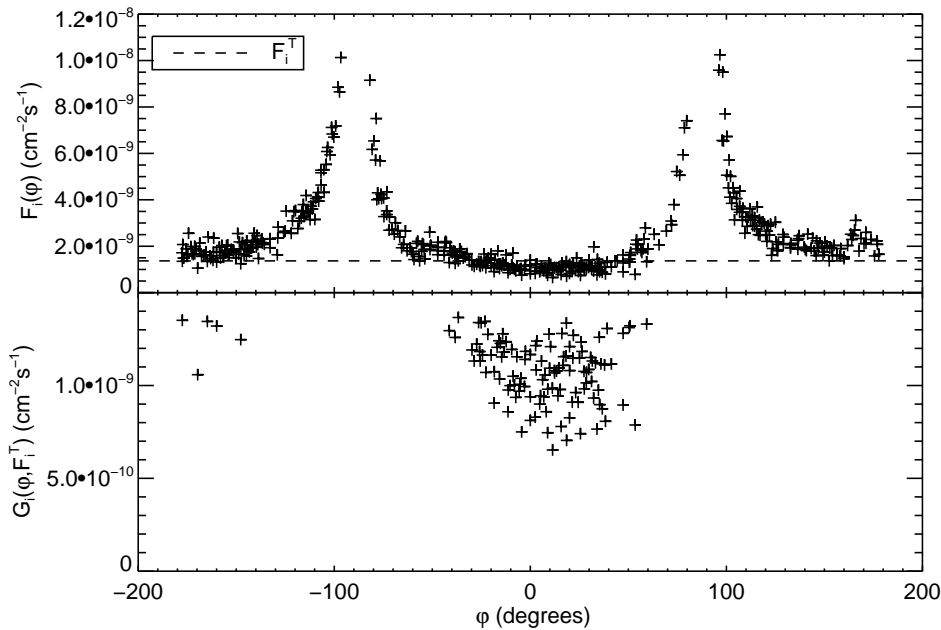}
\caption{An illustration of the terms defined in the text for the case
  of the data from the Fermi telescope. {\it Top panel:} The flux in a
  ring of constant $\psi_i$, $F_i(\phi_j)$, as a function of $\phi$,
  the angle transverse to $\psi$.  For the purposes of illustration,
  we have chosen a ring centered at $\psi_i = 15^{\circ}$ of width
  $6^{\circ}$.  The energy of the photons has been restricted to $1
  \gev < E < 100 \gev$.  The increase in flux at $\phi \approx \pm
  90^{\circ}$ is due to the galactic plane.  The dashed line
  illustrates the maximum flux that is consistent with the the data
  being i.i.d. as determined by the BDS test. {\it Bottom Panel:} The
  function $G_i(\phi_j,F^T_i)$ obtained by dropping all pixels with
  flux above the threshold $F^T_i$ in the top panel. This new set of
  fluxes is inconsistent with an i.i.d. sequence; a slightly lower
  value of $F^T_i$ would produce a $G_i$ that is consistent with an
  i.i.d. sequence. Therefore, the chosen value of $F^T$ is the
  strongest constraint on the i.i.d. flux in this ring.}
\label{fig:ring_phi_plot}
\end{figure}

Consider a ring on the sky of constant $\psi$ defined by $\psi_i -
\Delta \psi_i /2 < \psi < \psi_i + \Delta \psi_i /2$, where $\psi_i$
labels the central $\psi$ of the $i$th ring, and $\Delta \psi_i$ is
its angular width.  Since we have divided the sky into pixels, we
define $F_i(\phi_j)$ to be the flux in the $j$th pixel (labeled by its
azimuth angle $\phi_j$) of ring $i$.  We assume that the pixels have
been sorted in terms of $\phi_j$ so that, for example, pixels $j$ and
$j+1$ appear adjacent to each other on the sky; such sorting preserves
any non-i.i.d. component of the data.  The flux $F_i(\phi_j)$
receives contributions from an i.i.d. component due to the signal,
$F_{i,S}(\phi_j)$, and contributions from some possibly non-i.i.d.\,
components due to the backgrounds, $F_{i,B}(\phi_j)$ (see upper panel
of Fig.~\ref{fig:ring_phi_plot} for an illustration). Our goal is to
obtain an upper limit on the mean contribution from the signal.

Consider the fluxes in all pixels in the ring shown in
Fig.~\ref{fig:ring_phi_plot}. This sequence is clearly {\it not}
i.i.d. because of the peaks at $\phi\pm90^\circ$. That is, the flux in
the pixel at $\phi=90^\circ$ is clearly correlated with the flux in
nearby pixels. The physical reason for this is clear as well: these
values of $\phi$ correspond to the plane of the galaxy where
backgrounds are particularly large.  The mean flux of any
i.i.d. signal in this ring is clearly much smaller than the flux at
the peaks. Choosing as a constraint the mean flux in all pixels is
also not optimal as the plane of the Galaxy is distorting the
mean. Rather, we expect the constraint on the mean i.i.d. flux to be
at the level of the flux away from these peaks. One way to arrive at
this systematically is to:
\begin{itemize}
  \item Start with the observed distribution ($F_i(\phi_j)$) and note that
    it is not an i.i.d. sequence
  \item Create a flux threshold ($F_i^T$) and remove all pixels with
    flux above the threshold
  \item Test and see if the resulting (truncated) sequence
    ($G_i(\phi_j,F_i^T)$) is consistent with being i.i.d.
  \item If the truncated sequence is not consistent with an
    i.i.d. distribution, lower the threshold and repeat
  \item Once the truncated distribution is consistent with i.i.d., any
    i.i.d. component will generally have mean flux that is below the
    median of the truncated sequence, $F_i^{\rm{UL}} =
    \textrm{median}(G_i(\phi_j,F_i^T))$, so this median value becomes the
    upper limit on the i.i.d. flux in the ring (there are exceptions
    to this statement, an issue which we address below)
  \end{itemize}  

To determine whether a given $G_i(\phi_j, F_i^{T})$ is i.i.d., we use
the Brock, Dechert and Scheinkman (BDS) statistic \cite{Brock:1986}.
The BDS statistic tests the null hypothesis that a sequence is
i.i.d. by measuring the degree of spatial correlation in the sequence.
In essence, this is accomplished by searching for sub-sequences of
length $m$ that are significantly different from other $m$-long
sub-sequences in the data; the value of $m$ is referred to as the
`embedding dimension'.  The null hypothesis can be rejected if the BDS
statistic falls outside of some desired confidence interval; in our
analysis we use a $3\sigma$ confidence limit.  See
Ref.\cite{Cromwell:1994} for an introduction to the BDS statistic.  To
implement the BDS test we use a code made available in
Ref.~\cite{LeBaron:1997}.  The test itself depends on two parameters:
the maximum embedding dimension, $m_{max}$, and a parameter $\epsilon$
which effects how the discrepancy between different $m$-sequences is
measured.  Following the recommendations of Ref.~\cite{Brock:1991}, we
use $\epsilon =0.5\sigma$ where $\sigma$ is the standard deviation of
$G_i(\phi_j, F_i^{T})$ and $m_{max} = 5$; we find, however, that our
results are not very sensitive to the choices of these parameters. The
results of the BDS test can be easily checked by eye; any
non-i.i.d. behavior is obvious to the eye as spatial clumping.

Ideally, the limit we derive through this method will be significantly
lower than the mean flux in the ring, $\overline{F_i(\phi_j)}$,
because we have effectively removed the contributions to the ring that
are not i.i.d.  The upper panel of Fig.~\ref{fig:ring_phi_plot} shows
the initial $F_i(\phi_j)$ for a particular ring on the sky; the lower
panel shows $G_i(\phi_j, F_i^{T})$, where $F_i^{T}$ has been
determined using the BDS test.  

\subsection{Monte Carlo Testing}

There are several important qualifications to the above discussion.
First, it is possible to engineer pathological signals and backgrounds
such that the limit determined by the Ring Analysis method is actually
lower than the mean signal flux.  Any realistic astrophysical sources
are unlikely to have such pathological distributions, however.
Second, the point spread function (PSF) of the Fermi telescope at 1
GeV is comparable to the size of our pixels; one might worry that the
non-zero PSF could lead to correlations between pixels that would
invalidate the i.i.d. property of the signal.  Third, the statistical
literature recommends that the BDS test be used only on data sets that
are large, preferably with more than 500 elements \cite{Brock:1991}.
In our analysis, however, we apply the test to data sets with as few
as 50 elements, clearly pushing the limits of the BDS test. Finally,
if the signal is a smoothly varying function of $\psi$ then it will
not have a constant value in any ring of finite angular width since
different pixels at fixed $\psi_i$ integrate over the flux with
different weighting over $\psi$.  All of the above issues can be
addressed through the application of Monte Carlo tests.  By applying
the Ring Analysis technique to simulated data sets, we can determine
to what extent and under what conditions our constraints are valid.

Since we aim only to place an upper limit on the mean flux of an
i.i.d. component in the data, the relevant statistic for evaluating
the success of the Ring Analysis technique is the ratio of the
calculated upper limit on the i.i.d. flux, $F^{\rm{UL}}$, to the mean
i.i.d. flux, $\mu_{\rm{i.i.d.}}$.  Ideally, $F^{\rm{UL}}/\mu_{\rm{i.i.d.}}$ will
always be greater than one (so that our constraint is valid) but not
much greater than one (so that the constraint is as tight as
possible). We wish to characterize how this statistic varies as a
function of the mean i.i.d. flux.

We generate mock data sets by combining a simulated i.i.d. signal and
a simulated background.  These mock data sets are then analyzed using
the Ring Analysis technique and its performance is evaluated.  The
mock i.i.d. component is drawn randomly from a Poisson distribution
with a desired mean.  The random draws ensure that the signal is in
fact i.i.d. and the assumption of a Poisson distribution is justified
for most astrophysical sources.  For the background model, it makes
sense to use the observed data itself since the data is likely
background dominated.  However, since any i.i.d. component to the
background will increase $F^{\rm{UL}}/\mu_{\rm{i.i.d.}}$, we subtract
the determined i.i.d. flux from the observed background to generate
our background model.  This is the most conservative test of our
analysis: if the true background contains any i.i.d. component then
the true $F^{\rm{UL}}/\mu_{\rm{i.i.d.}}$ can only be larger than the
$F^{\rm{UL}}/\mu_{\rm{i.i.d.}}$ measured in the Monte Carlo trials.

One might worry that the point spread function (PSF) of the FGST could
lead to correlations between pixels that might disturb the
i.i.d. nature of the underlying signal.  The width of the FGST PSF is
a declining function of energy; at the lowest energy that we consider
(1 GeV), the 68\% containment angle is slightly less than 1 degree
(for normal incidence; the PSF broadens slightly when the incidence
angle is beyond about $50^{\circ}$).  Consequently, the width of the
PSF at the lowest energies considered is comparable to the size of our
$\sim 1^{\circ}$ pixels.  To account for effects of the PSF we
applying a Gaussian smoothing kernel to the mock signal with a
standard deviation of $0.5$ pixels.  Since the PSF actually gets
narrower at higher energies, the application of such a smoothing
kernel to all of the photon data is conservative; at high energies we
are overestimating the size of the PSF.

As mentioned above, we do not expect the signal in an angular ring of
finite width to be exactly constant even if the signal is azimuthally
symmetric since the signal is a smoothly varying function.  To account
for this possibility in our Monte Carlo tests we introduce a variation
in the mean signal flux that amounts to a $40\%$ decrease across the
ring.  This value is chosen because it is characteristic of the
variation in the smooth dark matter signal in which we are ultimately
interested.

\begin{figure}[htpb]
\includegraphics[scale = 0.7]{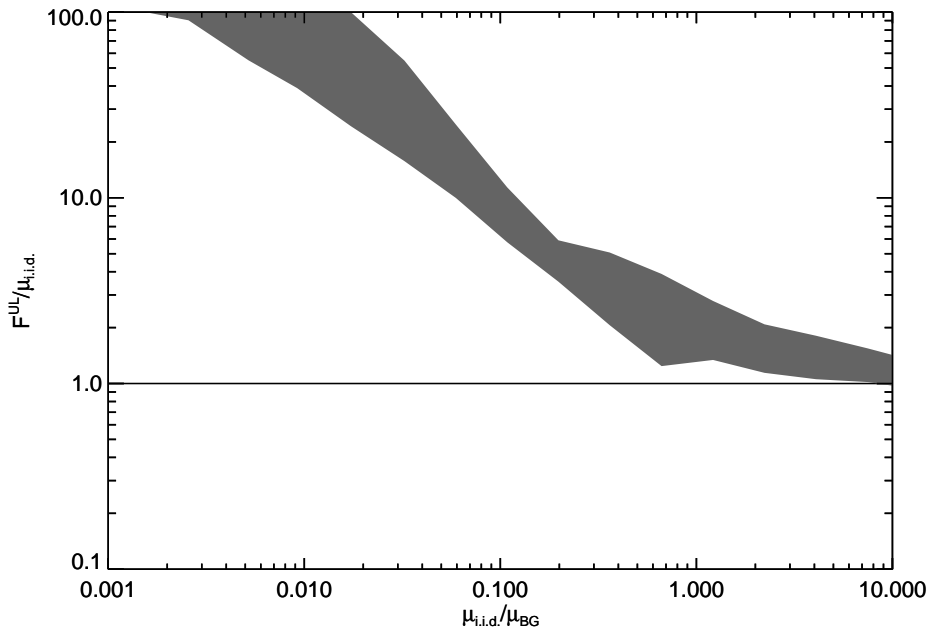}
\caption{Results of the Monte Carlo trials for evaluating the
  performance of the Ring Analysis.  The y-axis represents the ratio
  of the upper limit flux, $F^{\rm{UL}}$, determined using the ring
  analysis to the mean flux of the mock i.i.d. signal,
  $\mu_{\rm{i.i.d}}$; the x-axis represents the ratio of the mean flux
  of the mock i.i.d. signal to the mean flux of the mock
  background. The shaded band indicates the 95\% confidence interval
  of the Monte Carlo trials; 1000 trials were conducted for each value
  of $\mu_{\rm{i.i.d.}}$.  From the figure it is clear that the Ring
  Analysis is performing as hoped: $F^{\rm{UL}}/\mu_{\rm{i.i.d.}}$ is
  always larger than one (so that the constraint is valid) and it
  stays close to one over a fairly large range of $\mu_{\rm{i.i.d.}}$
  (so that the constraint is tight).}
\label{fig:threshold_bounds}
\end{figure}

Fig.~\ref{fig:threshold_bounds} presents the results of our Monte
Carlo tests of the Ring Analysis method.  The figure shows the
quantity $F^{\rm{UL}}/\mu_{\rm{i.i.d.}}$ for the mock data sets as a
function of $\mu_{\rm{i.i.d.}}/\mu_{BG}$, where $\mu_{BG}$ is the mean
background flux (the signal and background have been simulated as
described above with the effects of the PSF and the smoothly varying
signal included).  For each value of $\mu_{BG}$ we drew 1000
realizations of the i.i.d. signal.  The shaded region indicates the
95\% confidence interval for these 1000 mock data sets.
Fig.~\ref{fig:threshold_bounds} reveals that the Ring Analysis
technique is performing essentially as hoped:
$F^{\rm{UL}}/\mu_{\rm{i.i.d.}}$ is always larger than one and it stays
close to one over a fairly large range of $\mu_{\rm{i.i.d.}}$.  For
very low $\mu_{\rm{i.i.d.}}$ the signal makes essentially no
contribution to the data and the flux threshold therefore stays
constant; consequently $F^{\rm{UL}}/\mu_{\rm{i.i.d.}} \propto
1/\mu_{\rm{i.i.d.}}$ in this regime.

\subsection{Constraints on the i.i.d. Flux}

\begin{figure}[htpb]
\includegraphics[scale = 0.7]{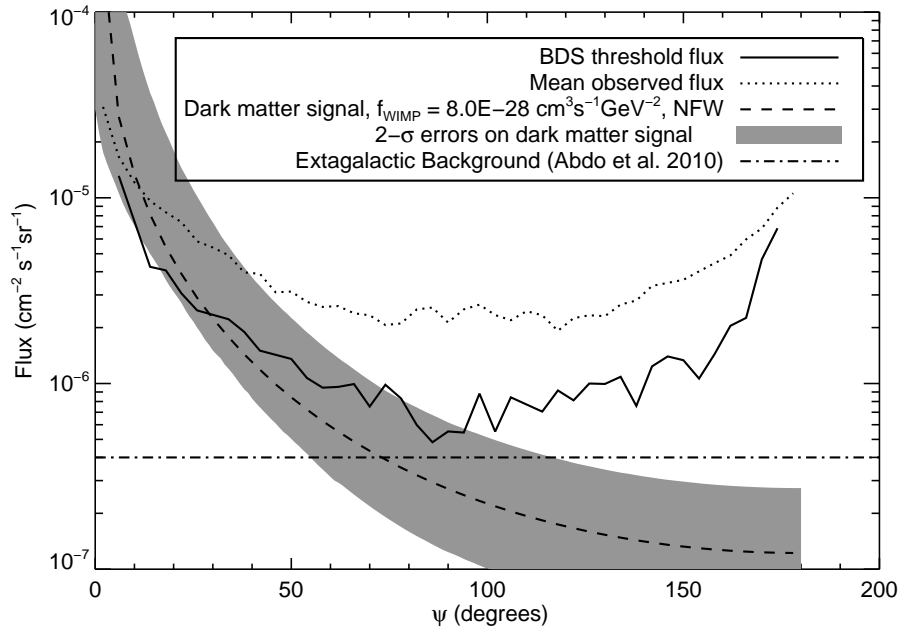}


\caption{Plot of the upper limit flux, $F^{\rm{UL}}_i$, in the $i$th
  ring (solid line) determined using the BDS test as a function of the
  angle from the Galactic Center, $\psi$, at which the ring lies.
  This threshold flux represents the maximum flux possible for any
  i.i.d. component in that ring.  The dotted curve shows the mean flux
  in the $i$th ring for comparison.  The dashed curve represents the
  expected signal from the smooth dark matter in our galaxy assuming
  $f_{\mathrm{WIMP}}=8\times 10^{-28} \textrm{cm}^3 \textrm{s}^{-1}
  \textrm{GeV}^{-2}$ and a canonical halo model; the shaded region
  represents the errors on the expected dark matter signal when
  uncertainties in the halo properties are taken into account.  The
  dash-dotted line corresponds to the value of the extragalactic flux
  determined by the analysis of Ref.~\cite{Abdo:2010}.  We have
  restricted the data in the plot to $1 \gev < E < 100 \gev$ and have
  used 45 rings of equal angular width in $\psi$ across the sky.}
\label{fig:ring_psi_plot}
\end{figure}

Fig.~\ref{fig:ring_psi_plot} shows the upper limits on an
i.i.d. component in the Fermi data derived using the Ring Analysis
technique as a function of the angle $\psi$ from the galactic
center. As one moves away from the Galactic Center (i.e. towards $\psi
= 90^{\circ}$), the limit becomes tighter simply because the flux is
lower; beyond $\psi = 90^{\circ}$ the limit becomes weaker again as we
look toward the galactic anticenter.  In generating
Fig.~\ref{fig:ring_psi_plot} we divided the sky into 45 rings of equal
width in $\psi$.  The BDS threshold curve is shown only for those
rings for which the test could be conducted on more than 50 pixels in
accordance with the results of our Monte Carlo simulations.  As can be
seen from the figure, the BDS threshold is significantly more
constraining than the mean flux in the ring.

As a consistency check on our results, we compare the minimum
threshold flux in Fig.~\ref{fig:ring_psi_plot} (which occurs near
$\psi = 90^{\circ}$ as expected since this high-latitude region is
least contaminated by the Galaxy) to the value of the extragalactic
\gray background determined by the analysis of the Fermi Collaboration
in Ref.~\cite{Abdo:2010} (see the dash-dotted line in
Fig.\,\ref{fig:ring_psi_plot}).  Since any isotropic, smooth
background is necessarily i.i.d. we expect our minimum threshold flux
to be at least as large as the measured isotropic background.  We do
not expect the minimum threshold flux to be much larger than the
isotropic background, however, because the isotropic background
dominates at $\psi \approx 90^{\circ}$.  As expected, our minimum
threshold flux of roughly $5\times10^{-7} \rm{cm}^{-2} \rm{s}^{-1}
\rm{sr}^{-1}$ is slightly larger than the value of $\sim4\times10^{-7}
\rm{cm}^{-2} \rm{s}^{-1} \rm{sr}^{-1}$ obtained by
Ref.~\cite{Abdo:2010} for the extragalactic \gray background flux over
the same energy range ($1 \gev < E < 100 \gev$).

Also shown in Fig.~\ref{fig:ring_psi_plot} is the flux predicted from
the smooth dark matter component for a particular particle mass and
cross section (\fwimp is defined in the next section) and a smooth
Navarro-Frenk-White (NFW)~\cite{Navarro:1996} distribution with
canonical values for the total halo mass and concentration. The shaded
region (described in the next section) reflects the uncertainties in
the dark matter distribution.  Roughly, then, these values of the mass
and cross section are ruled out by the Ring Analysis. In the next
section, we will project this constraint on to the mass--cross section
plane and propagate the uncertainties in the dark matter distribution
to this plane. Here, we note that the most stringent limit comes from
the ring\footnote{Note that our analysis is restricted to $\psi\ge
  7^\circ$. While the constraint could in principle be improved by
  moving to smaller $\psi$, the small number of pixels at such $\psi$
  means that the BDS test loses much of its statistical power.  We
  have confirmed with Monte Carlo tests that the constraints placed by
  the Ring Analysis tests become invalid in this regime. For the
  purposes of constraining the dark matter signal,
  Fig.~\ref{fig:ring_psi_plot} suggests that the innermost region will
  not tighten the constraints.} with $\psi=15^\circ$; this
model-independent limit is:
\begin{equation}
F_{\rm{i.i.d.}}(\psi=15^\circ)\le 4.5\times 10^{-6} \rm{cm}^{-2} \rm{s}^{-1} \rm{sr}^{-1}.
\eql{fluxlim}
\end{equation}
The fact that this analysis identifies the annulus at $\psi=15^\circ$
as the most constraining is consistent with the arguments of
Ref.~\cite{Stoehr:2003hf}.


Fig.~\ref{fig:ring_psi_plot} also shows the mean flux in each
ring. The mean flux is always larger than the i.i.d. upper limit; in
the most constraining ring at $\psi=15^\circ$, the i.i.d. limit is a
factor of 2 below the mean flux. This factor of two illustrates the
power of the i.i.d. analysis: the limit obtained from the mean in the
ring is contaminated by flux near the Galactic plane, contamination
that is removed by the Ring Analysis introduced here.

The main advantage of the Ring Analysis is that it makes no
assumptions about the source or properties of the backgrounds to the
dark matter signal.  This is a significant advantage because
uncertainties in backgrounds are the main limitation for constraining
dark matter with \gray observations.  We have assumed only that the
signal in each pixel is drawn from a distribution that is invariant
under rotations around the Galactic Center.  Of course, the cost of
assuming little is that the limit that can be placed using this
technique is comparatively weak.  For instance, since any isotropic
backgrounds present in the data will also meet these assumptions,
their presence will make our determined limit worse.  Furthermore, we
note that this technique is not well suited for actually detecting
dark matter, but rather should be viewed as a way of placing upper
limits on the dark matter signal.

\section{Constraints on Dark Matter}
\label{sec:application}

\subsection{The Dark Matter Signal}

The Milky Way is believed to exist within a roughly spherical halo of
dark matter \cite{Zaritsky:1999}.  Consequently, the annihilation
signal from galactic dark matter is expected to have azimuthal
symmetry around the line connecting us to the Galactic Center.  The
dark matter constituting the halo is in turn thought to exist in two
forms: a smooth component and a clumped component termed {\it
  subhalos} \cite{Diemand:2007}.  Almost by definition, a smooth dark
matter component will have an annihilation signal for which variations
from the mean are uncorrelated.  Therefore, the annihilation signal
from smooth dark matter is a perfect candidate for the Ring Analysis
developed in the previous section.  Here we ignore potential signals
from annihilations occurring in subhalos, as well as a possible
cosmological signal from dark matter annihilations. Therefore, the
constraint -- on only the smooth Galactic component -- is
conservative.

In order to turn the flux limit derived by the Ring Analysis into
constraints on the dark matter particle properties, we must develop a
model for the expected annihilation flux from smooth Galactic dark
matter.  The observed flux of photons due to such annihilations along
a given line of sight is
\begin{eqnarray}
\label{eq:smoothflux}
F(\psi) = \frac{f_{\rm{WIMP}} J(\psi) }{4 \pi}
\end{eqnarray}
where the first factor on the right
\begin{equation}
\label{eq:fwimp}
f_{\rm{WIMP}} \equiv
\frac{N_{\gamma} \langle\sigma v \rangle} {M_{\chi}^2}
\end{equation}  
depends on the particle physics properties of the dark matter: mass
$M_\chi$, velocity-weighted annihilation cross section $\langle \sigma
v \rangle$, and photon counts per annihilation
\begin{eqnarray}
\label{eq:ngamma}
N_{\gamma} = \int_{E_{min}}^{E_{max}} \frac{dN}{dE} dE
\end{eqnarray}
within some desired band $E_{min} < E < E_{max}$. The second factor in
\ec{smoothflux} 
\begin{equation}
J(\psi) \equiv \int dl d\Omega \rho^2(l, \psi) 
\end{equation}
depends on the distribution of the dark matter halo, with
$\rho(l,\psi)$ the smooth dark matter density at line of sight
distance $l$ and zenith angle $\psi$.  We have assumed that $\rho$ is
spherically symmetric about the Galactic Center so that the observed
flux can be written as a function of $\psi$ only but will revisit this
assumption later.  A given profile and therefore a value of $J$
translates the constraints on the i.i.d. flux in the previous section
into a constraint on \fwimp.

A canonical $J$ emerges from assuming an NFW profile with a scale
radius of 20 kpc and local dark matter density of $\rho_0 = 0.43
\gev \rm{cm}^{-3}$.  This leads to the $J(\psi)$ depicted as the
dashed line in Fig.~\ref{fig:ring_psi_plot} for the given value of
\fwimp.

\subsection{Uncertainties in the Dark Matter Profile}

The dark matter profile is not constrained very tightly by
observations, and this uncertainty propagates to the constraints on
the particle physics properties. We address this in two
ways here, the first applies to all constraints on the smooth halo and
the second is specific to our assumption that the dark matter
distribution is spherical.

To allow for freedom in the dark matter profile, we piggy-back on the
analysis of Ref.~\cite{Widrow:2008}.  They used nine sets of
observational data to constrain the dark matter profile in our
Galaxy. They assumed a density of the form
\begin{equation}
\label{eq:galdensity}
\rho(r) = \frac{2^{2-\gamma}\sigma_h^2}{4 \pi a_h^2
  G}\frac{1}{(r/a_h)(1+r/a_h)^{3-\gamma}}
\end{equation}
where $r$ is the distance from the Galactic Center, $a_h$ is the scale
radius of the dark matter halo, $\sigma_h$ is a scale velocity, and
$\gamma$ is a parameter which effects the behavior of the central
density cusp, with $\gamma = 1$ corresponding to the usual NFW form. A
given set of these free parameters translates into a set of values for
$J(\psi)$ assuming the distance between the Earth and the Galactic
Center to be 8.5 kpc.

\begin{figure}[htbp]
\includegraphics[scale = 0.7]{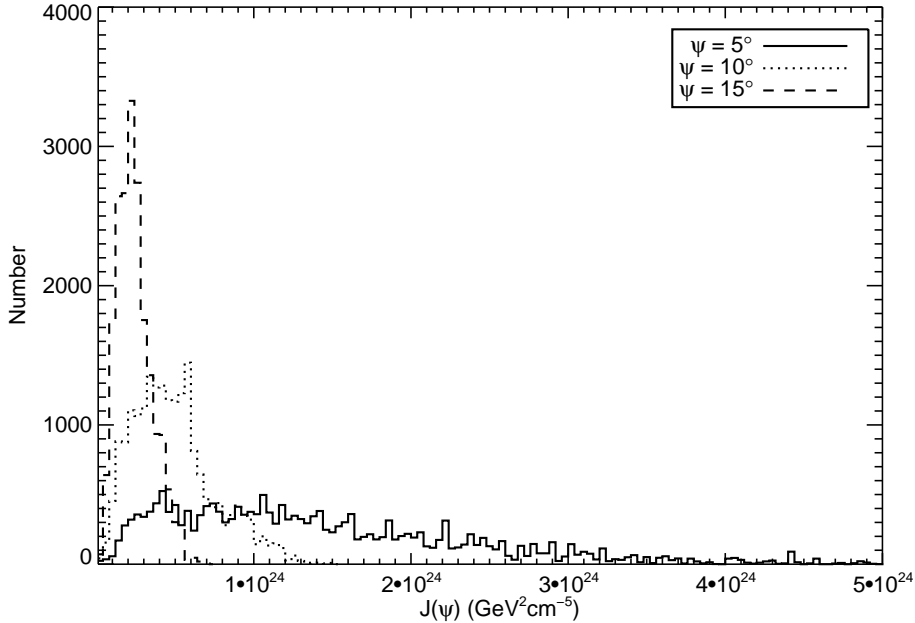}
\caption{Histogram of values of $J(\psi) = \int dl \rho^2(\ell,\psi)$
  corresponding to models in the MCMC chains provided by
  \cite{Widrow:2008}.  We show the histograms for $\psi = 10^{\circ}$
  and $\psi = 15^{\circ}$ since these angles provide the strongest
  constraint on dark matter annihilations. The broad histogram for
  $\psi=5^\circ$ illustrates the well-known statement that the
  uncertainty in the flux increases towards the Galactic Center.}
\label{fig:integral_hist}
\end{figure}

The allowed values of the three profile parameters (and therefore
$J(\psi)$) are contained in Monte Carlo Markov Chains run in
Ref.~\cite{Widrow:2008}. They have kindly provided us with those
chains, and Fig.~\ref{fig:integral_hist} projects these on to
$J(\psi)$ for three different values of $\psi$.  It is clear from this
figure that the width of the distributions increases rapidly with
decreasing $\psi$.

The widths of the distributions show in Fig.~\ref{fig:integral_hist}
can be propagated directly into error bars on the expected signal for
dark matter in Fig.~\ref{fig:ring_psi_plot}.  The shaded region in
Fig.~\ref{fig:ring_psi_plot} represents the 95\% confidence interval
for the allowed dark matter signal for a particular choice of \fwimp.
The value we have chosen for \fwimp is illustrative in the sense that
it is roughly the lowest value that is excluded by the Ring Analysis
limit.

\begin{figure}[htbp]
\includegraphics[scale = 0.7]{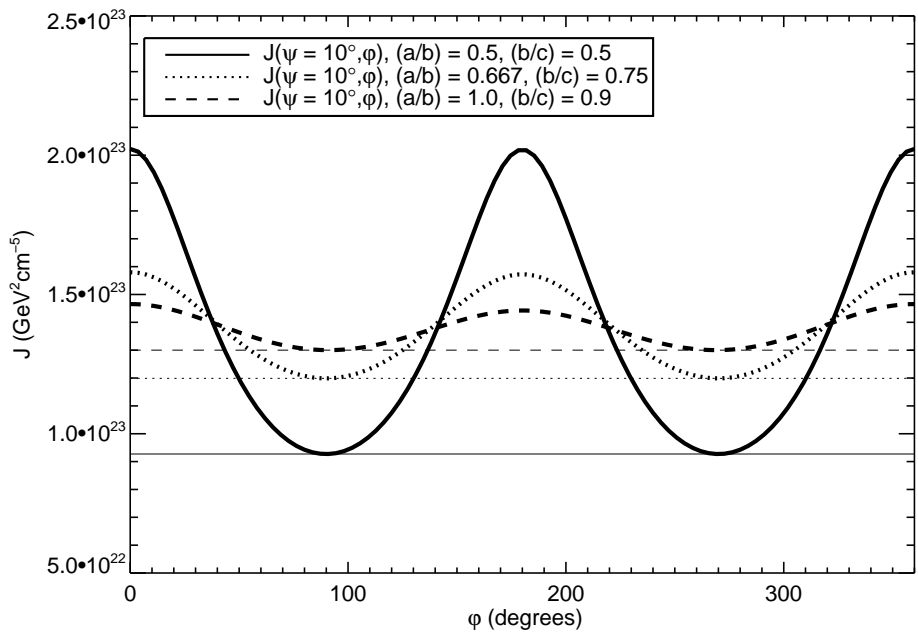}
\caption{The value of $J(\psi=10^{\circ}, \phi)$ along different lines of sight
  for three model triaxial galaxies (thick lines).  The minor,
  intermediate and major axes are labelled by $a$, $b$ and $c$
  respectively.  Also plotted are the corresponding levels of the
  i.i.d. component of the signal (thin lines).  Following the results
  of Ref.~\cite{Law:2009} we have assumed that the minor and major axes of
  the halo lie in the galactic plane and that the line connecting the
  sun to the Galactic Center is offset from the minor axis of the halo
  by $15^{\circ}$.}
\label{fig:triaxial_j}
\end{figure}

Our analysis so far has assumed that the signal from the smooth
component of the dark matter is spherically distributed around the
Galactic Center.  In fact, the true shape of the Milky Way's dark
matter halo may be triaxial \cite{Law:2009}.  If this is the case,
then the dark matter signal is not uniform in rings of constant $\psi$
on the sky; instead, the signal in such a ring will be an oscillating
function of the azimuth angle $\phi$ (the thick curves in
Fig.~\ref{fig:triaxial_j}).  As a result, the total signal in a given
ring will no longer be i.i.d. (i.e., it is not {\it identical}).
However, there will still be some component of the signal that
\textit{is} i.i.d.; the flux of this component is given by the minimum
flux of the dark matter signal in the ring (the thin horizontal lines in
Fig.~\ref{fig:triaxial_j}).  Since the triaxiality of the halo
effectively decreases the magnitude of the i.i.d. signal, the stated
lower bound is too aggressive.  It is therefore important to quantify
how the triaxiality of the halo effects the level of the
i.i.d. signal.

In order to estimate the magnitude of this effect, we calculate
\begin{equation}
J(\psi, \phi) \equiv \int dl d\Omega \rho^2(l, \psi, \phi)
\end{equation}
along different lines of sight in model triaxial galaxies.  $J(\psi,
\phi)$ is the analogue of $J(\psi)$ for a dark matter halo that is not
assumed to be spherically symmetric.  Observations in the Milky Way by
Ref.~\cite{Law:2009} and simulations of disk galaxies by
Ref.~\cite{Bailin:2005} suggest that one of the axes of the inner halo
(which is also the region that dominates our exclusion limit) is
aligned with the rotation axis of the galactic disk.  Following the
results of Ref.~\cite{Law:2009}, we assume that the minor and major
axes of the halo lie in the galactic plane; we find, however, that
this choice of alignment does not have a significant impact on our
results.  We assume a density profile of the form
Eq. \ref{eq:galdensity} with the replacement
\begin{eqnarray}
\label{eq:triaxialradius}
r^2 = (x/c)^2 + (y/a)^2 + (z/b)^2
\end{eqnarray}
where the $x$, $y$ and $z$ axis are aligned with the minor, major and
intermediate axes of the halo respectively.  Strictly speaking,
Refs.~\cite{Jing:2002,Allgood:2006} have found that the axis ratios
$a/b$ and $b/c$ do not remain constant throughout the halo; rather,
they decrease slightly (i.e. the ellipsoid becomes more elongated)
towards the center of the halo.  However, since our exclusion limit is
dominated by a small range of distances from the Galactic Center we
feel justified in approximating the halo by ellipsoids with constant
axis ratios and alignments throughout the galaxy as per
Eq.~\ref{eq:triaxialradius}.

Fig.~\ref{fig:triaxial_j} shows $J(\psi = 10^{\circ}, \phi)$ along
different lines of sight in model galaxies with several values of the
axis ratios (thick lines).  In generating this plot we have used
$\sigma_h = 270 \textrm{km/s}$, $a_h =5.9 \textrm{ kpc}$ and $\gamma =
0.028$; in a spherical halo these values yield a J($\psi =
10^{\circ}$) which is at the fifth percentile of all those calculated
from the Markov chains of Ref.~\cite{Widrow:2008}.  Since the
uncertainty in our exclusion limit is dominated by uncertainty in the
halo model, this parameter choice is therefore appropriate for our
$95\%$ exclusion limit on the dark matter properties.  We plot only
those lines of sight with $\psi = 10^{\circ}$ because this is the
angular range relevant to our exclusion limit.  All of the curves have
been normalized so that they have the same mean as the corresponding
curve for a spherical halo; since the total amount of dark matter in
the inner halo is strongly constrained (by measurements of galactic
rotation curves, for instance), we expect this choice of normalization
to be reasonably accurate.

To proceed further, we need to know the axis ratios of the isodensity
surfaces throughout the Milky Way halo.  Ref.~\cite{Law:2009} determined
values for the axis ratios of some of the isovelocity surfaces in the
Milky Way.  While these values could in principle be converted to axis
ratios for the isodensity surfaces given a density model, the large
uncertainties associated with these measurements lead us to take a
hopefully more robust approach.  Using numerical simulations,
Refs.~\cite{Jing:2002,Allgood:2006} have determined probability
distributions for the axis ratios of dark matter halos.  We run Monte
Carlo simulations based on these results in order to quantify the
extent of the variation in the dark matter signal due to triaxiality.

We calculate $J(\psi, \phi)$ for halos with axis ratios drawn from the
probability distributions of Eqs. 17 and 18 in Ref.~\cite{Jing:2002}.  We
adjust these distributions slightly to account for the fact that they
are calculated at a distance from the Galactic Center, $r_{2500}$,
given by $\rho(r_{2500})/\rho_{\mathrm{crit}} = 2500$ (note that
$r_{2500}$ is defined in terms of the local halo density and not the
mean interior density) while we are interested in the distributions
much closer to the Galactic Center.  Using Eqs. 6 and 7 of
Ref.~\cite{Jing:2002} we scale the axis ratios so that they correspond to a
distance of $\sim 1.5 \textrm{ kpc}$ from the Galactic Center, roughly
the innermost distance that effects our constraint.  This corresponds
to decreasing $a/c$ and $b/c$ by roughly 30\% and 10\% respectively.
The resultant distributions are roughly consistent with the results of
Ref.~\cite{Allgood:2006}.  Finally, we assume that the line connecting the
sun to the Galactic Center lies in the plane of the minor and major
axes of the halo but with a random orientation in that
plane.

\begin{figure}[htbp]
\includegraphics[scale = 0.7]{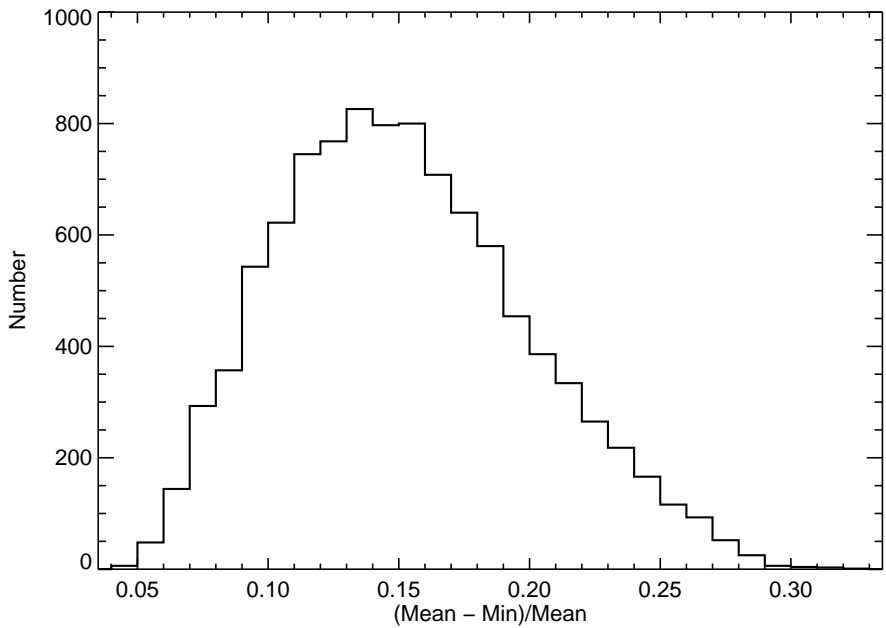}
\caption{Distribution of the percentage variation (mean to minimum) of
  the signal from smooth dark matter at $\psi = 10^{\circ}$ in 10000
  realizations of our Monte Carlo modeling of triaxial halos.  Axis
  ratios have been drawn from the probability distributions of
  Ref.~\cite{Jing:2002} (scaled according to the prescription
  described in the text).  The line connecting the sun to the Galactic
  Center is assumed to lie in the plane of the minor and major axes of
  the halo, but with a random orientation in that plane.}
\label{fig:oscillation_hist}
\end{figure}

Fig.~\ref{fig:triaxial_j} shows the flux as a function of azimuthal
angle $\phi$ in a given ring for three different sets of
$(a,b,c)$. The key take-away from these plots is the difference
between the mean flux (i.e. the mean value of
$J(\psi=10^{\circ},\phi)$ for the thick curves in
Fig.~\ref{fig:triaxial_j}) and the minimum flux (i.e. the thin curves
in Fig.~\ref{fig:triaxial_j}) in the ring.  This difference is the
amount by which we have been implicitly {\it overestimating} the
i.i.d. signal.  Dividing the difference by the mean flux then provides
an estimate of the relative over-estimation, the amount by which we
should loosen the i.i.d. constraint.  Fig.~\ref{fig:oscillation_hist}
shows the distribution of this fractional difference (also at $\psi =
10^{\circ}$) for 10000 Monte Carlo realizations of the halo axis
ratios drawn from the probability distributions of
Ref.~\cite{Jing:2002}.  It is clear from this figure that error
introduced by assuming a spherical halo is typically no more than
$25\%$.  Since the uncertainty introduced into our exclusion limit by
the uncertainties in $\sigma_h$, $a_h$ and $\gamma$ is much greater
than this, and since the degree of triaxiality in our own halo is not
very well constrained, we choose to ignore this source of uncertainty
in our exclusion plots.

We conclude this section by mentioning that the presence of baryons
may have a non-negligible impact on the shape of the dark matter halo,
particularly in the innermost region \cite{Springel:2004}.  The
effects of baryons have not been included in the simulations of
Ref.~\cite{Jing:2002} nor Ref.~\cite{Allgood:2006}.  While there may
be a disturbance due to baryons near disk, the effect of such a
disturbance on our exclusion limit is likely negligible since the
limit is dominated by the region away from the disk where the
backgrounds are lowest.

\subsection{Constraints on Dark Matter}

The previous discussion has constrained \fwimp over the energy range
1\gev to 100\gev.  We find that our final constraint over this energy
range is
\begin{equation}
f_{\rm{WIMP}} \le 5.8\times 10^{-28} {\rm cm}^3\,{\rm s}^{-1}\, {\rm GeV}^{-2}
\end{equation}
In order to turn the limits on \fwimp into constraints on the dark
matter particle properties we must assume an annihilation channel for
the dark matter so that $N_{\gamma}$ (Eq.~\ref{eq:ngamma}) can be
calculated.  The constraint on \fwimp and the value of $N_{\gamma}$
can then be transformed into a constraint in the $M_{\chi}$-$\langle
\sigma v \rangle$ parameter space using Eq.~\ref{eq:fwimp}.  


The choice of the energy range that we impose on our analysis can have
a significant impact on the constraints that we place in the
$M_{\chi}$-$\langle \sigma v \rangle$ plane.  For $E \ll M_{\chi}$ the
annihilation photon spectrum is relatively flat compared to that of
the backgrounds, so increasing $E_{\mathrm{min}}$ in this regime
effectively increases the size of the signal relative to the total
flux, thereby improving the constraint on the dark matter properties.
As $E$ approaches $M_{\chi}$, the spectrum of annihilation photons
falls off very quickly so increasing $E_{\mathrm{min}}$ in this regime
causes the constraint to become very weak.  We expect, then, that the
optimal constraint on the dark matter comes from choosing
$E_{\mathrm{min}}$ to be some fixed fraction of $M_{\chi}$.  Rather
than enforcing the optimal $E_{\mathrm{min}}$ for each $M_{\chi}$
exactly (which would require dividing the data into many more energy
bins than the 29 that we employ), we allow $E_{\mathrm{min}}$ to vary
freely for each $M_{\chi}$ that we consider.  We then choose the value
of $E_{\mathrm{min}}$ that maximizes our constraint on the dark
matter.  Random fluctuations in the strength of the constraint induced
by varying $E_{\mathrm{min}}$ are small, so choosing
$E_{\mathrm{min}}$ to maximize the constraint does not decrease the
statistical significance of our result.  Varying $E_{\mathrm{min}}$ is
our one minimal use of the energy information for the photons; by
making our analysis essentially independent of detailed spectral
information we make our constraints more robust.

\begin{figure}[htbp]
\begin{tabular}{cc}
\includegraphics[scale = .7]{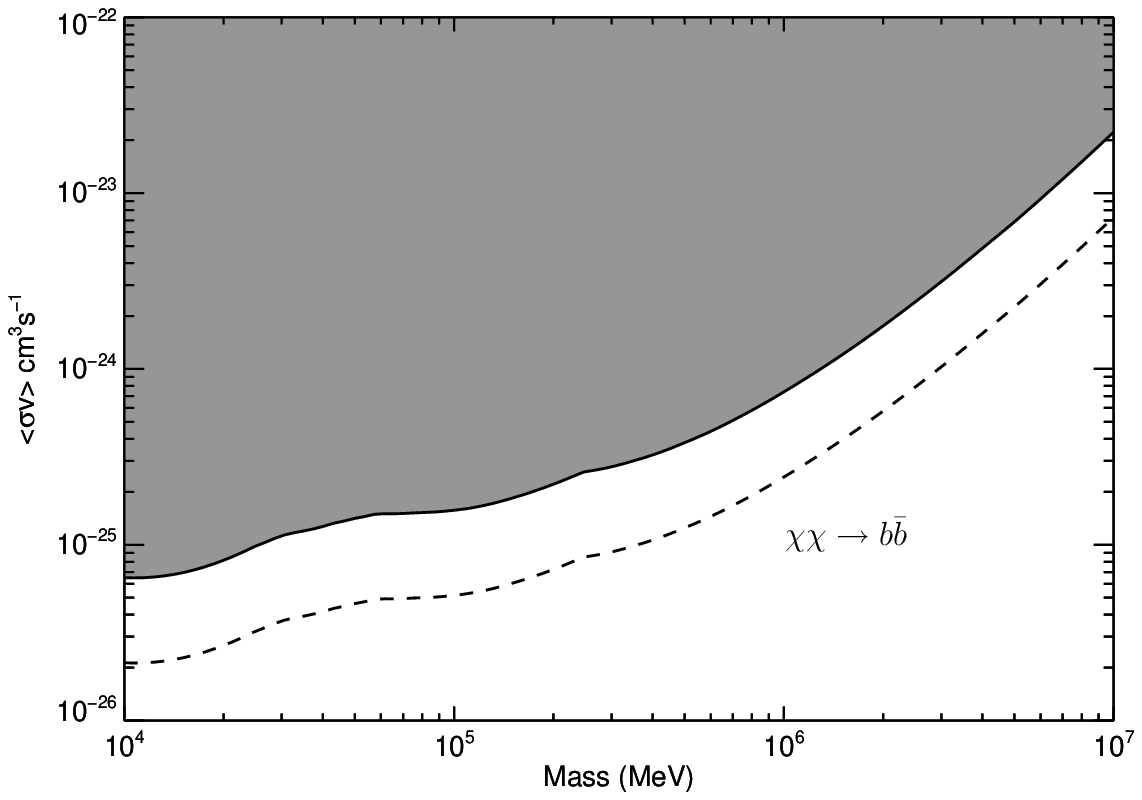} & \includegraphics[scale=.7]{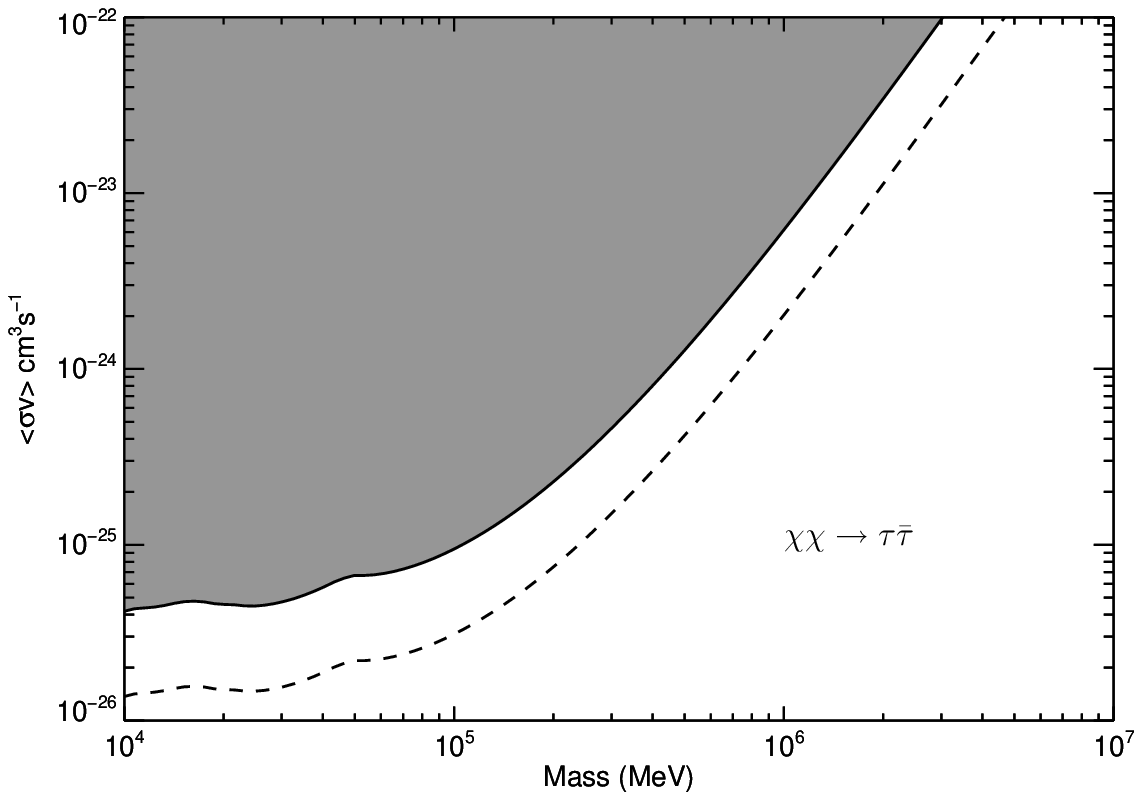}  
\end{tabular}
\caption{Exclusion plots in the $\langle\sigma v \rangle$-$M_{\chi}$
  plane generated by the Ring Analysis for different annihilation
  channels of the dark matter.  The solid region represent the 95\%
  exclusion range with respect to the uncertainties in the properties
  of the smooth halo.  The dashed line represents the boundary of the
  excluded region when the mean value of $J(\psi)$ from the Monte
  Carlo chains of Ref.~\cite{Widrow:2008} is used to calculate the
  expected dark matter signal.  The left panel shows the exclusion
  limit for neutralinos, $\chi$, that annihilated to $b\bar b$; the
  right panel corresponds to neutralinos annihilating to
  $\tau\bar\tau$.}
\label{fig:ring_exclusion_channel}
\end{figure}

Fig.~\ref{fig:ring_exclusion_channel} shows how the constraints that
we place on \fwimp in different energy bins with the Ring Analysis are
translated into constraints in the $M_{\chi}$-$\langle\sigma v
\rangle$ plane for different assumed annihilation channels (each of
which correspond to a different $N_\gamma$).  We consider two possible
annihilation channels: $\chi\chi \rightarrow b \bar b$ and $\chi\chi
\rightarrow \tau \bar \tau$, where $\chi$ is a neutralino.
$N_{\gamma}$ is calculated for these different channels over a mass
range of 10 GeV to 10 TeV using the DarkSUSY\footnote{P. Gondolo,
  J. Edsjö, P. Ullio, L. Bergström, M. Schelke, E.A. Baltz,
  T. Bringmann and G. Duda, http://www.darksusy.org}
package~\cite{Gondolo}. Since our technique places a limit on
$f_{\rm{WIMP}} = N_{\gamma} \langle\sigma v \rangle / M_{\chi}^2$ we
expect the limit on \sigv to go roughly as $M_{\chi}^2$.  The
flattening of the limit at low mass is due to a decrease in
$N_{\gamma}$ as more and more annihilation photons fall outside of the
energy limits of our analysis.

The dashed curves in Fig.~\ref{fig:ring_exclusion_channel} show what
the limit would be if the dark matter profile, as quantified by
$J(\psi)$, produced the mean annihilation signal, and there was no
uncertainty in the profile. The uncertainty in dark matter profile
then loosens the constraint on the annihilation cross section by more
than a factor of 2.


The limits presented in Fig.~\ref{fig:ring_exclusion_channel} are
comparable to the corresponding plots from Ref. \cite{Zaharijas:2010},
which also considers the annihilation signal from the smooth galactic
dark matter component, but which makes a stronger set of assumptions
about the signal and backgrounds.  Our limits are complementary in the
sense that they are almost entirely independent of any assumptions
about the diffuse \gray background.  We also find that our constraints
are comparable to those obtained from an analysis of extragalactic
dark matter annihilations by Ref. \cite{Abdo:2010dm}.  As evidenced in
Fig.~5 of that work, the constraints derived from extragalactic
annihilations are subject to very large uncertainties in the magnitude
of the dark matter signal.  While uncertainties in the signal from
smooth galactic dark matter are also important, they are not nearly as
large.  Finally, our limits are also comparable to (and competitive
with) those derived from an analysis of the annihilation signal from
dwarf spheroidal galaxies by Ref. \cite{Abdo:2010ds}.

\section{Summary}
\label{sec:summary}

We have developed a technique for constraining the presence of a
smooth, azimuthally symmetric signal on the sky.  We showed with Monte
Carlo simulations that the technique is robust (i.e. the limits
derived from the technique are never below the mean signal flux).  By
applying the technique to data from the Fermi Gamma-Ray Space
Telescope we derived a constraint (\ec{fluxlim}) on the presence of
any smooth \gray signal that is symmetric with respect to rotations
about the axis connecting us to the Galactic Center.  When combined
with a model for the signal due to annihilations of the smooth dark
matter component in our galaxy, these limits allow us to place
constraints on the dark matter particle mass and cross section.  While
our limits are slightly weaker than other recent results, they have
the advantage of making essentially no assumptions about the
backgrounds to the dark matter signal.  This is a significant
advantage because the uncertainties in models of \gray backgrounds are
large and often unknown.

There are several ways these limits can be improved. Tighter
constraints on the dark matter profile would reduce the uncertainty in
$J(\psi)$, which currently degrades the ultimate limit by more than a
factor of 2. Including other sources of signal, in particular the
contribution from extra-galactic halos, would improve the signal, but
in most models, this contribution is smaller than that from the
Galactic halo. A deeper understanding of the backgrounds could also be
used in conjunction with this method to push the limits down further.

\vspace{1cm} 

{\it Acknowledgments} We are very grateful to Larry Widrow for
providing us with the chains from Ref.~\cite{Widrow:2008} and to
Andrey Kravtsov for his guidance on the properties of the Galactic
halo. This work has been supported by the US Department of Energy,
including grant DE-FG02-95ER40896, and by National Science Foundation
Grant AST-0908072.

\end{document}